# Wi-Fi, WiMax and WCDMA
# A comparative study based on Channel Impairments and Equalization method used

Rabindranath Bera[1], Sanjib Sil[2], Sourav Dhar[1] and  Subir K. Sarkar[3]

[1] Sikkim Manipal Institte of  Technology, Sikkim Manipal University, majitar, rangpo, East Sikkim 737132.
[2] Techno India College of Technology, Megacity, Newtown Kolkata 700156
[3] Dept. of Electronics & Telecommunication engineering, Jadavpur University, kolkata 700032

Email: r.bera@rediffmail.com

**Abstract:**
In this paper we describe the channel impairments and equalization methods currently  used in WiFi, WiMax and WCDMA. After a review of channel model for Intelligent Transportation System (ITS), we proposed an equalization method which will be useful for the estimation of strong multipath channel at a high velocity.

**Key Words:** RAKE, Adaptive Equalizer, Correlator.

## I. INTRODUCTION

The use of vehicle-to-vehicle (V2V) communication will be an integral part of intelligent transport systems (ITSs) [1], and work on ITS has been growing substantially in recent years[2]. Some obvious benefits of ITS are its ability to improve road safety; weather [3], making commuters aware of current traffic [4], and road conditions in real time; easing "bottlenecks" at toll booths, thus saving time and money for commuters and government; and turning long journeys into times for family activities by enabling the flow of multimedia between different traveling cars. The list of possible applications [5] is long. The scope of V2V communications is not limited to a fixed number of *a priori* specified vehicles and can hence be extrapolated to numerous vehicles via the concept of vehicular mobile ad hoc networks. Vehicular ad hoc networks are important since they remove the dependence on cellular networks for communication between vehicles. Public safety applications may also employ V2V communications [6]. The information flowing between vehicles will likely be multimedia: data, images, video, and voice. To popularize use of V2V communication devices among consumers, it is essential to highlight other possible applications such as the provision of entertainment services or other leisure related data. Vehicular ad hoc networks are also gaining importance for V2V communications, because they allow vehicles in close proximity to communicate without depending on other systems, e.g., the cellular network.

Inter symbol interference (ISI) is the major reason for erroneous reception of information in high speed wireless broad band communications. The use of adaptive equalization can reduce  the effect of ISI significantly [7]. Equalization virtually produces an ideal channel through which voice, data and video can pass trough and can be received without error. This paper presents a comparative study of the equalization techniques used in Wi-Fi, WiMax and in WCDMA based 3G mobile communication which are used in ITS applications. Wi-Fi and WiMax use orthogonal frequency division multiplexing (OFDM) technology while WCDMA uses direct sequence spread spectrum (DSSS) technology. Moreover variations in the channel due to variation of physical environment are different in these three wireless technologies. Hence the methods of equalization are also different. Wi-Fi is used for short range (less than 500 yards) communications; on the contrary WiMax and WCDMA are used for long range communications.

In this work, simulation is done, using MATLAB/ SIMULINK, to obtain the response of these three systems for different multipath channel scenario and observed how the equalization method is used for solving the problems arose due to the channel impairments.

## II. CHANNEL IN ITS

ITS application basically involves vehicle to vehicle (V2V) and vehicle to infrastructure (V2I) communications [8]. The development of the future Vehicle-to-Vehicle (V2V) and Vehicle-to-Infrastructure (V2I) communications systems imposes strong radio channel management challenges due to their decentralized nature and the strict Quality of Service (QoS) requirements of traffic safety applications. Fig.1 shows a pictorial representation of road condition and respective multipath scenario [9]. A large amount of previous research on the V2V channel topic has pertained to millimeter wave (mm wave) bands, e.g., [10], which usually use highly directional antennas to circumvent the large propagation path losses encountered at those frequencies. Some recent works have also gathered measured data in other bands, including the 2.4 GHz unlicensed band. Examples of these works include [11], which reported results for Doppler spread, and [12], which characterized attenuations in short range indoor mobile-to-mobile settings. Analytical studies have also been done for this type of channel, without regard to carrier frequency, including [13]-[15]. Most mm wave applications are short-range applications and it is difficult to provide an "always connected" communication link between vehicles using such point-point links except in some highway situations. Hence it would be beneficial to have a "broader range" network than can be sustained using only these directional point-point links. This would enable multiple vehicles to be in contact irrespective of the obstacles and traffic conditions among them, at larger distances than are achievable with the point-to-point links. This broader network could operate as a mobile LAN, and transceivers in this kind of V2V communication system will encounter channels that are much different from those for both mmwave and terrestrial cellular scenarios. Some reasons for this are that transmitter (Tx) and receiver (Rx) and some significant reflectors/scatterers are all mobile, the omnidirectional antennas for both Tx and Rx are at relatively low elevations, and because of the physical environment dynamics, the channel is statistically non-stationary.

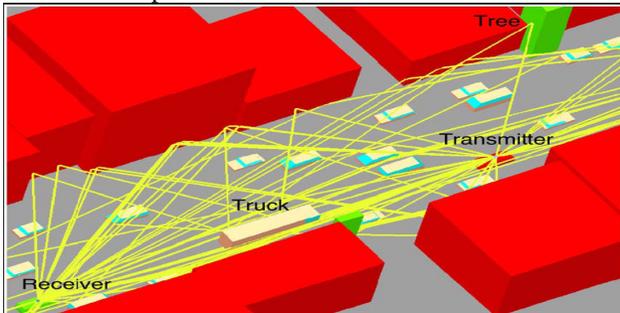

Fig.1. 50 strongest propagation paths

### III. WCDMA

*1. Channel Impairments:*

Matlab channel visualization tools are utilized for channel impulse response for WCDMA and it is observed that impulse response is broadened due to multipath effect. ISI could be reduced by narrowing the impulse response by the use of signal processing. Fig. 2 shows the IR response of the ideal channel having band width of 5 MHz and no multipath. Therefore it is desired that sampling points must be located at 200ns apart. Now the delay vector is changed (i.e. distance of the scatterer is changed) by a small amount in order to visualize the IR response. It is found that the main lobe of the impulse response of the system is broadened by an amount of 5.90% (Fig. 3) in both the cases.

*2. Equalization in WCDMA*

Rake is used for equalization in WCDMA [16]. Radio propagation in the land mobile channel is characterized by multiple reflections, diffractions and attenuation of the signal energy. These are caused by natural obstacles such as buildings, hills, and so on, resulting in so-called multipath propagation. To improve the signal to noise ratio (SNR) at the receiver RAKE receiver can combine multi-path components. Rake provides a separate correlation receiver for each of the multi-path signals and Multi-path components are practically uncorrelated when their relative propagation delays exceeds one chip period. It uses a multi-path diversity principle that means it rakes the energy from the multi-path propagated signal components.

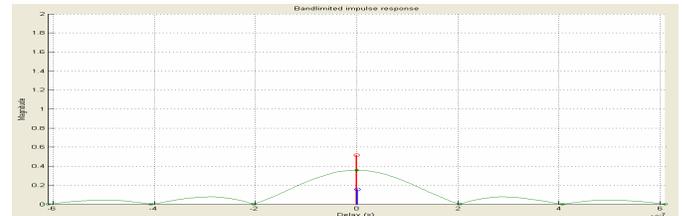

Fig 2: Impulse Response of Ideal channel

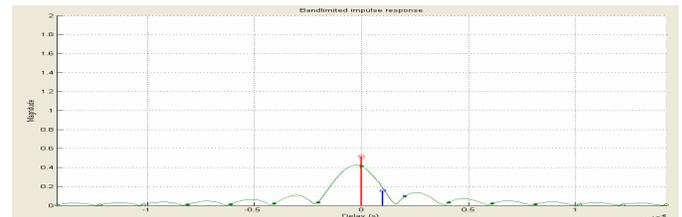

Fig 3: Impulse Response of channel after placing the scatterer at a distant position.

RAKE receiver utilizes multiple correlators to separately detect strongest multi-path components. Each correlator detects a time-shifted version of the original transmission, and each finger correlates to a portion of the signal, which is delayed by at least one chip in time from the other fingers. The outputs of each correlator are weighted to provide better estimate of the transmitted signal than is provided by a single component. Strong multipath signals will appear as peaks, as illustrated in the path searcher block in Fig.4. The searcher sets a peak detection threshold to avoid false triggering and selects the best four peaks. The associated time delays for each of the peaks are forwarded to the fingers' fine finger tracking, code generator, and delay equalizer. The path searcher is usually computationally more intensive than the Rake because of this multiple search process.

MMSE correlator based rake receiver algorithm replaces the conventional correlator [17] in each finger with MMSE correlator, thus mitigating multi-user access interference while reducing computational complexity. Another method suggests increasing the number of correlator within a wider bandwidth to improve performance and achieve gains on the expense of modest increase in complexity .Thus, adding more correlator will require a repetition of the finger structure as required as depicted in Fig. 5, and a reconstruction of the combining part of the receiver. Here we have successfully implemented multiple-correlator with 4-Finger for mitigating multipath effects through MATLAB simulation (Fig.6). TABLE I shows some of the results obtained after combining the output of all the RAKE fingers as well as energy received by individual fingers. It is evident from the results that RAKE fingers are capable of extracting and combining the energy from the multipath components.

### IV. WiFi

*1. Channel Impairments:*

WiFi channel impairments are simulated by keeping a scatterer (e.g. building) between the transmitter and receiver. The channel is visualized using channel visualization tool of MATLAB. The scatterer is kept at different places (Fig.7). Fig. 8 shows the channel impulse response (IR) and frequency response (FR) with single scatterer. Red line indicates the LOS component where as blue indicates NLOS. Fig. 9 shows that if the scatterer is very close to LOS then its strength is minimum and then its strength increases as it is moved further and attains a maximum value. Now if scatterer is moved further from LOS, then its strength decreases.

TABLE I

| Magnitude of Tx signal (in dB) | Fing-1 (in dB) | Fing-2 (in dB) | Fing-3 (in dB) | Fing-4 (in dB) | Total Magnitude measured at Rx side (in dB) |
|---|---|---|---|---|---|
| 1 | 1.100 | 1.062 | 0.996 | 1.152 | 4.313 |
| 3 | 3.060 | 3.007 | 2.936 | 3.090 | 12.10 |
| 6 | 6.025 | 5.960 | 5.888 | 6.038 | 23.91 |
| 7 | 7.018 | 6.951 | 6.878 | 7.028 | 27.87 |
| 10 | 10.00 | 9.932 | 9.857 | 10.01 | 39.81 |
| 15 | 15.00 | 14.92 | 14.82 | 14.99 | 59.75 |
| 20 | 19.99 | 19.92 | 19.84 | 19.99 | 79.74 |
| 25 | 24.99 | 24.91 | 24.84 | 24.99 | 99.73 |

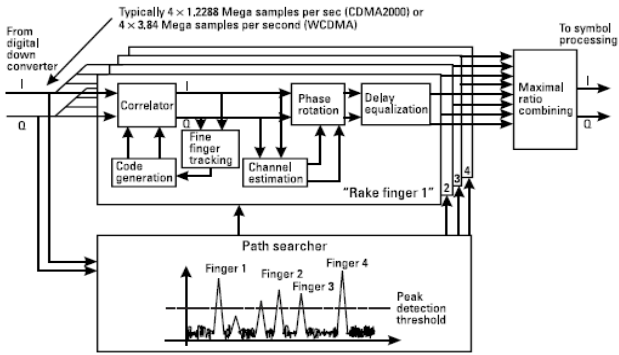

Fig. 4: Rake receiver for multipath rejection.

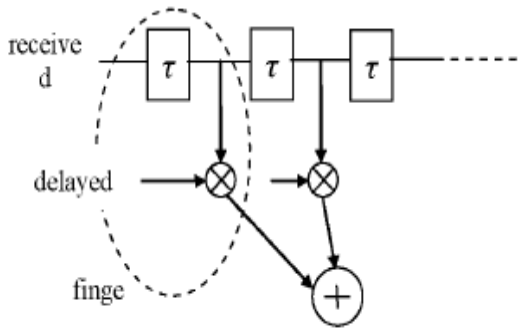

Fig. 5: More correlator could be constructed in a repetitive way.

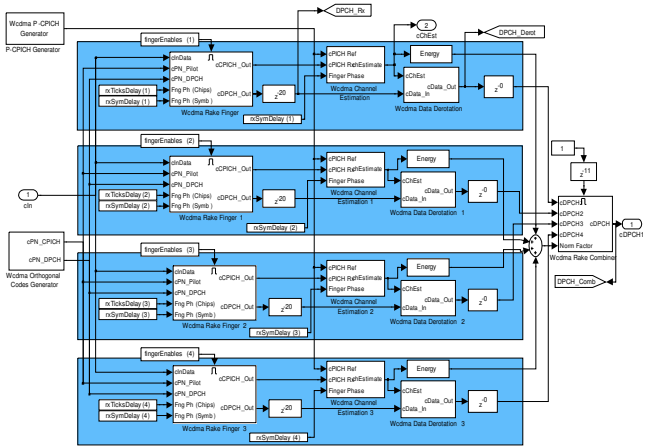

Fig. 6: Simulation of RAKE receiver for WCDMA

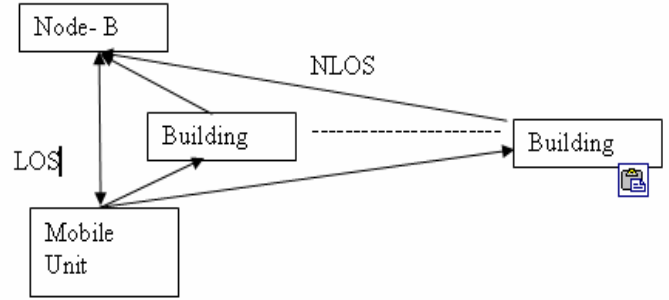

Fig.7: Simulation model for channel impairment measurement

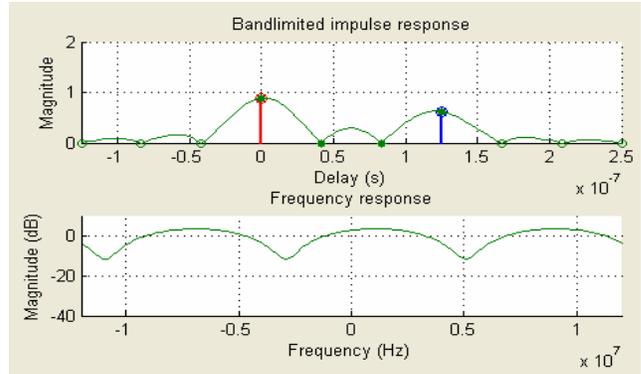

Fig. 8: Channel impulse response and frequency response with single scatterer

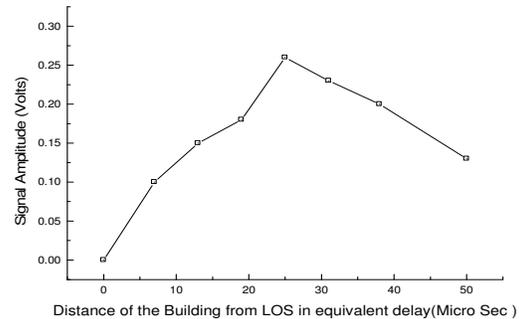

Fig.9: Effect of scatterer position on received signal

*2. Equalization in WiFi*

WiFi uses adaptive equalizer to mitigate channel impairments. Equalizer can be used as time & frequency domain filters. An equalizer in which the coefficients are dynamically configured in order to approximate the optimum results is an adaptive equalizer. Several convergence algorithms exist for updating these coefficients, and each represents a different trade-off between the computational complexity required to compute the taps at each step and the speed with which the coefficients come close to the optimum values, i.e. the convergence speed. Hence, convergence speed is one of the most important and deciding factors for the consideration of different types of algorithm used in adaptive equalizer. In all cases, the convergence algorithms require knowledge of the current error at the output of the equalizer, and appropriate means must be provided. It is clear that the correct signals are not known at the receiver in general, as this would require knowledge of the transmitted sequence. The usual approach is depicted in Fig 10 [7].

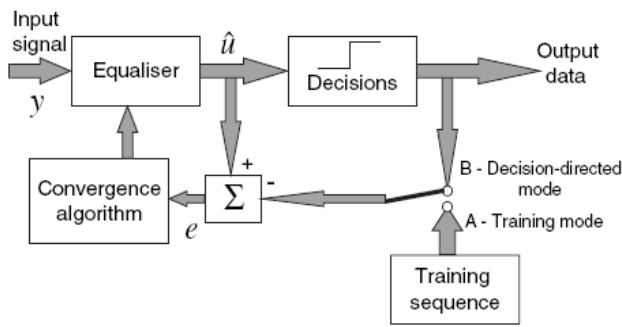

Fig. 10: Structure of an adaptive equalizer

The whole working of the adaptive equalizer is mainly divided into two parts:
→ first is the Training Mode for known sequence code,
→ second is the Decision-Directed Mode for unknown transmitted data.

Depending on the accuracy of the equalizer coefficients (which is approximated during first mode of operation) and the noise level, the decisions may sometimes be incorrect and this ultimately depends upon the Algorithm followed in different approach. This approach mainly takes care of ISI plus noise signal.

## V. WiMAX

*1. Channel Impairments:*

Channel estimation and equalization schemes for broadband wireless network are an active area for recent research. WiMAX (Worldwide Interoperability for Microwave Access) is a new release standard for this technology which still facing real challenge for low complexity and efficient implementation. WiMAX supports non-line-of-sight (NLOS) environment with high data rate transmission and high mobility up to 125 Km/hr [18]. It provides very high data throughput over long distance in a point-to multipoint and line of sight (LOS) or non-line of sight (NLOS) environments. WiMAX can provide seamless wireless services up to 20 or 30 miles away from the base station. The IEEE 802.16 group subsequently produced 802.16a, which include NLOS applications in the 2GHz–11GHz band, using an orthogonal frequency division multiplexing (OFDM)- based physical layer. Further revisions resulted in a new standard in 2004, called IEEE 802.16- 2004, which replaced all prior versions and formed the basis for the first WiMAX solution.

Fig 11 (a,b) shows received signal constellation and power spectrum before and after equalization.

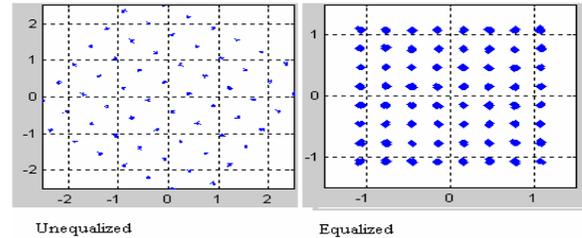

Fig. 11 a. Rx signal Constellation

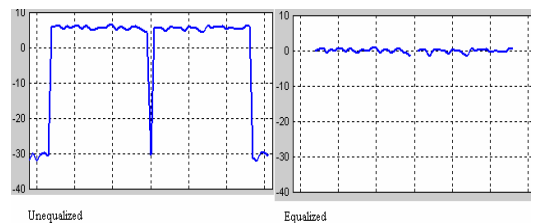

Fig. 11 b. Rx signal Rx Power spectrum

These early WiMAX solutions based on IEEE 802.16-2004 targeted fixed applications often it is referred to as fixed WiMAX [19]. OFDM technique is widely adopted in those systems due to it's robustness against Multipath fading and simpler equalization scheme. In most of applications, for retaining the orthogonality of subcarriers and overcome intersymbol interference (ISI), a cyclic prefix (CP) is inserted instead of simply inserting guard interval. If the maximum delay of the Multipath channel does not exceed the CP length, the OFDM system would be ISI free by removing the guarding interval. For WiMAX systems, its delay spread is typically over several microseconds which are longer than the guarding interval. Therefore, it is very challenging to maintain the system BER performance for non-line-of-sight (NLOS) channels at high data rate transmission also for mobile WiMAX **Doppler effect degrades system performance**. Fig. 12 shows the channel impairment with and without Doppler effect.

*2. Equalization in WiMax*

Least mean-square (LMS) equalizer is basically used for channel estimation in WiMax [18]. Fig. 13 shows received signal constellation before and after equalization. Here signal constellation is stabilized after equalization but not as good as in WiFi.

## VI. CONCLUSION

The three systems considered here are useful for different channel conditions. Thus different kind of equalizations are used to encounter the channel impairments and this leads to a proper corrective method. WiFi and Wimax use OFDM technology.

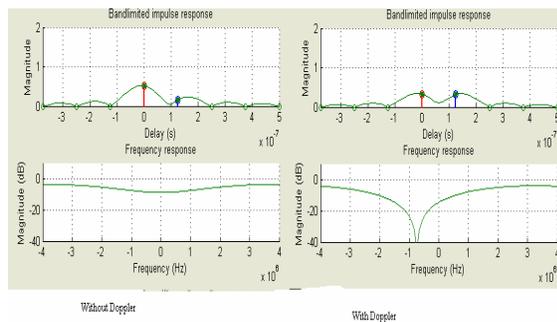

Fig. 12: Channel IR and FR with and without Doppler effect in WiMax

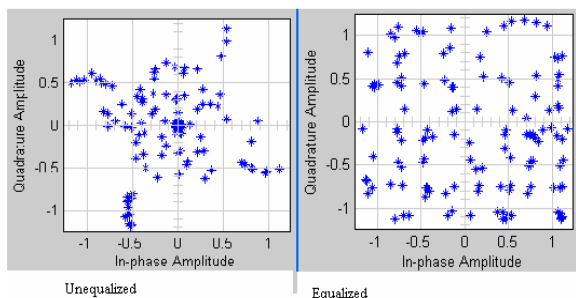

Fig. 13: Rx signal Constellation

If the maximum delay of the Multipath channel does not exceed the CP length, the OFDM system would be ISI free by removing the guarding interval. But at high speed V2V communications often delay spread is typically over several microseconds which are longer than the guarding interval. Also high Doppler velocity making channel estimation even more difficult. Fig. 14 shows a generalized cell structure for wireless communication. Position 'B' of Fig.14 indicates that the mobile unit is at cell boundary. Rake receiver used by WCDMA will treat the scattered signal as well as weaker signal received from either of the Base Stations (BS) as multipath signal and will adjust the correlator accordingly. Thus the effective SNR is more in WCDMA compared to adaptive (LMS) equalizers used in WiFi and WiMax.. Moreover, coded technology leads to soft handover which ensures data integrity in motion.

The above discussion leads the authors to conclude that the use of multiple correlator equalizer or RAKE is a must to mitigate multipath problems arise during ITS application.

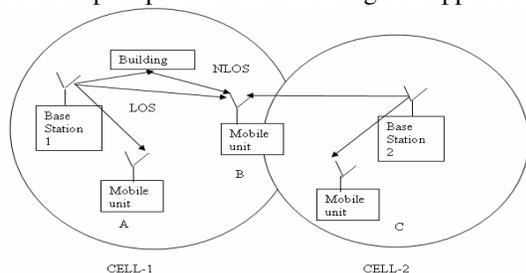

Fig. 14 : Generalized cell structure of WiFi, WiMax and WCDMA

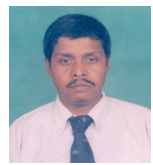

**Rabindra Nath Bera**: Born in 1958 at Kolaghat , West Bengal, INDIA. Received his B. Tech, M. Tech & Ph.D (Tech) from the Institute of Radiophysics & Electronics, The University of Calcutta, in the year 1982,1985 & 1997 respectively. Currently working as Dean (R&D) and Head of the Deparment, Electronics & Communication Engineering, Sikkim Manipal University,Sikkim, Microwave/ Millimeter wave based Broadband Wireless Mobile Communication and Remote Sensing are the area of specializations.